\documentclass[12pt]{article}
\usepackage{graphics}
\usepackage{axodraw}
\usepackage{epsfig}
\newcommand{\Py}{{\sc Pythia}}

\newcommand{\Ar}{{\sc Ariadne}}

\newlength{\captivewidth}
\newcommand{\captive}[1]{\rule{5mm}{0mm}%
\begin{minipage}{\captivewidth}%
\caption[small]{#1}\end{minipage}}
\def\Journal#1#2#3#4{{#1}{\bf #2} (#3) #4}

\def\PLB{{\it Phys. Lett.~}{\bf  B}}

\def\ZPC{{\it Z. Phys.~}{\bf C}}

\def\EPJC{{\it Eur. Phys. J.~}{\bf C}}

\begin{document}
\begin{flushright}
  LU TP 02-45 \\
  December 2002
\end{flushright}
\vspace{2cm}
\begin{center}
  \Large
  {\bf Recent Developments in the Lund Model\footnote{The article is compiled by S.Mohanty and F.S\"oderberg from the transparencies of the talk presented by the Late Prof. Bo Andersson.}} \\
  \normalsize \vspace{2mm}
  Bo Andersson  \\
  Sandipan Mohanty  \\
  Fredrik S\"oderberg \\

  Department of Theoretical Physics, Lund University, \\
  S\"olvegatan 14A, S-223 62 Lund, Sweden
\end{center}
\vspace{1cm}
\noindent  {\bf   Abstract}  A   brief  introduction  to   the  String
Fragmentation Model and its  consequences is presented. We discuss the
fragmentation of a general multi-gluon  string and show that it can be
formulated as  the production of  a set of ''plaquettes''  between the
hadronic curve and the  directrix. We also discuss certain interesting
scaling  properties  of the  partonic  states  obtained from  standard
parton  shower algorithms  like those  in {\Py}  and {\Ar},  which are
communicated  to   the  final  state  hadrons   obtained  through  our
hadronisation procedure.

\section{The Area Law and its Consequences}
In the Lund  String Model the massless relativistic  string is used to
model the  QCD field between coloured  objects, \cite{BA,BA1,BA2}. The
string has a  constant tension $\kappa$, which gives  rise to a linear
potential.  The  endpoints of the  string are identified as  the quark
and the anti-quark.   The endpoints will stretch the  string when they
move apart  and therefore lose energy-momentum.  Because  of this they
will sooner  or later  have to reverse  their direction of  motion and
again gain  energy-momentum from  the string. The  motion of  a simple
string system  with a massless  quark and anti-quark at  the endpoints
will therefore be the so-called yoyo-mode, \cite{BA,BA1,BA2}.  We will
now describe this string state in more detail.

In   order    to   be   more    specific,   let   us    consider   the
$e^+e^-$-annihilation process  in Fig.\ref{area}.  We  assume that the
quark and the anti-quark are produced at a single space-time point and
that  they start  to move  apart. This  means that  there will  be one
original  string  spanned  between   the  quark  and  the  anti-quark,
cf. Fig.\ref{area}. Gluons are  interpreted as internal excitations of
the string and they will complicate the string motion dramatically. In
this section  we will only consider string-states  without gluons, and
we will  come back  to multi-gluon strings  in the next  section.  The
final state  hadrons observed in high  energy $e^+e^-$-collisions stem
from the  breakup of the force  field. New quark  anti-quark pairs are
produced from the  field. This means that the  original string will be
split into  smaller string pieces  because the quarks  and anti-quarks
are identified as string endpoints in  the string Model. We are now going
to describe this break-up process in more detail.

A   typical   Lund  String   break-up   process   can   be  found   in
Fig.\ref{area}. A set of new $q \bar{q}$-pairs are produced at certain
space-time points (called  vertices). One can note the following 
things in this figure:

\begin{figure}[t]
\begin{center}
\epsfig{file=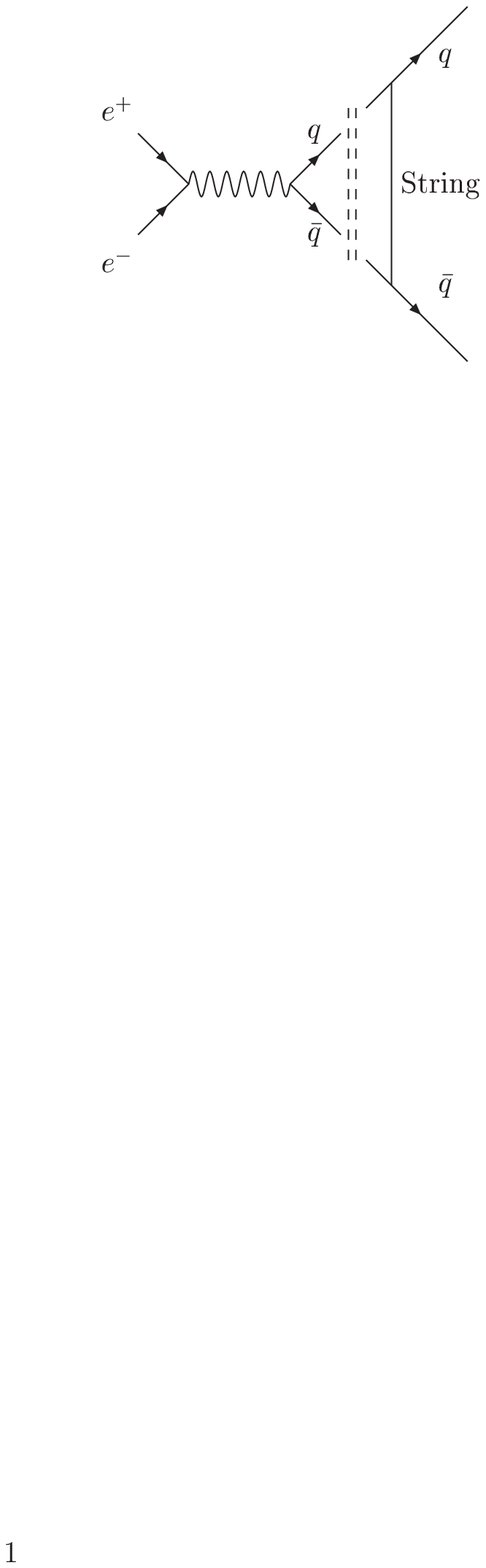,width=6.0cm}
\epsfig{file=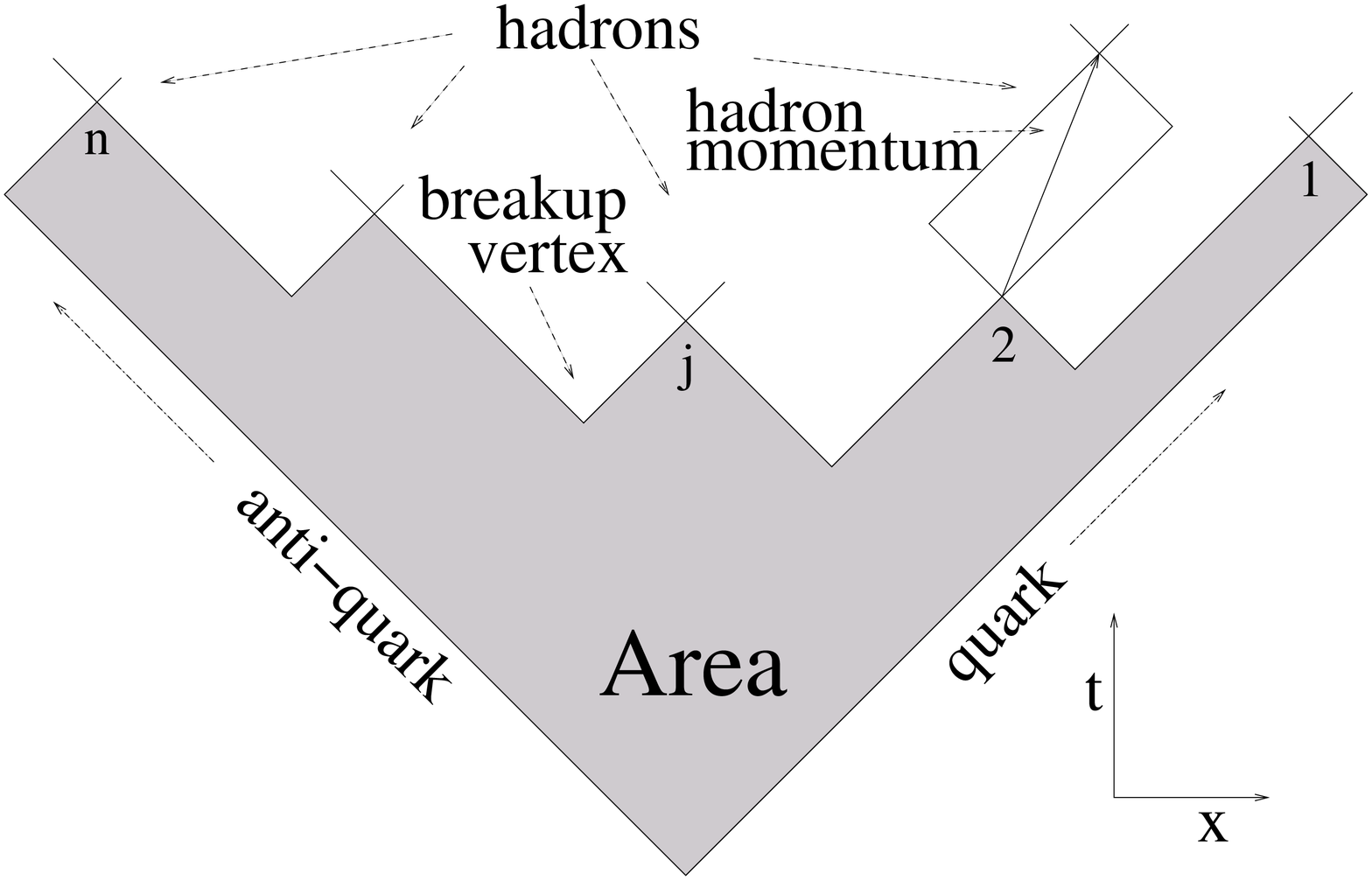,width=6.0cm}
\caption{The figure shows a typical Lund String break-up.
\label{area}}
\end{center}
\end{figure}

\begin{itemize}
\item  There is  no field  between a  $q \bar{q}$-pair  produced  at a
single vertex. A string force field is always confining because it has a fixed energy per unit length and the force field vanishes at the end-point charges.
\item  The  hadrons must  have  positive  momenta  which implies,  cf.
Fig.\ref{area}, that the vertices  must be space-like separated.  This
means  that  no invariant  time-ordering  is  possible.   There is  no
principle ''first'' vertex, they are  all ''equal''. In the Lund Model,
it will always be the slowest particles that are produced first in any frame.
\end{itemize}
\begin{figure}
\begin{center}
\epsfig{figure=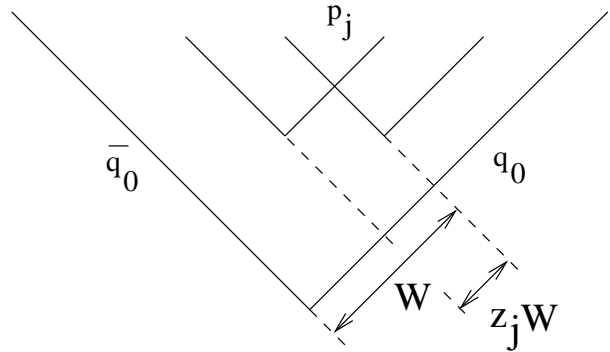, width=8.0cm }
\end{center}
\caption{Two adjacent vertices and a hadron with energy-momentum $p_j$ produced in between.}
\label{step}
\end{figure}
A convenient ordering prescription  is rank-ordering. Two hadrons that
share the same  vertex will be defined to have  adjacent rank.  If one
starts  to count  from the  original quark  end, this  means  that the
hadron which  contains this  quark and an  anti-quark from  a break-up
vertex will have rank 1, cf. Fig.\ref{area}.  The break-up process can
be described as a set of steps along the positive light-cone momentum,
where each hadron  takes a fraction $z_j$ of  the remaining light-cone
momentum, cf. Fig.\ref{step}.

\begin{figure}[!h]
\begin{center}
\epsfig{figure=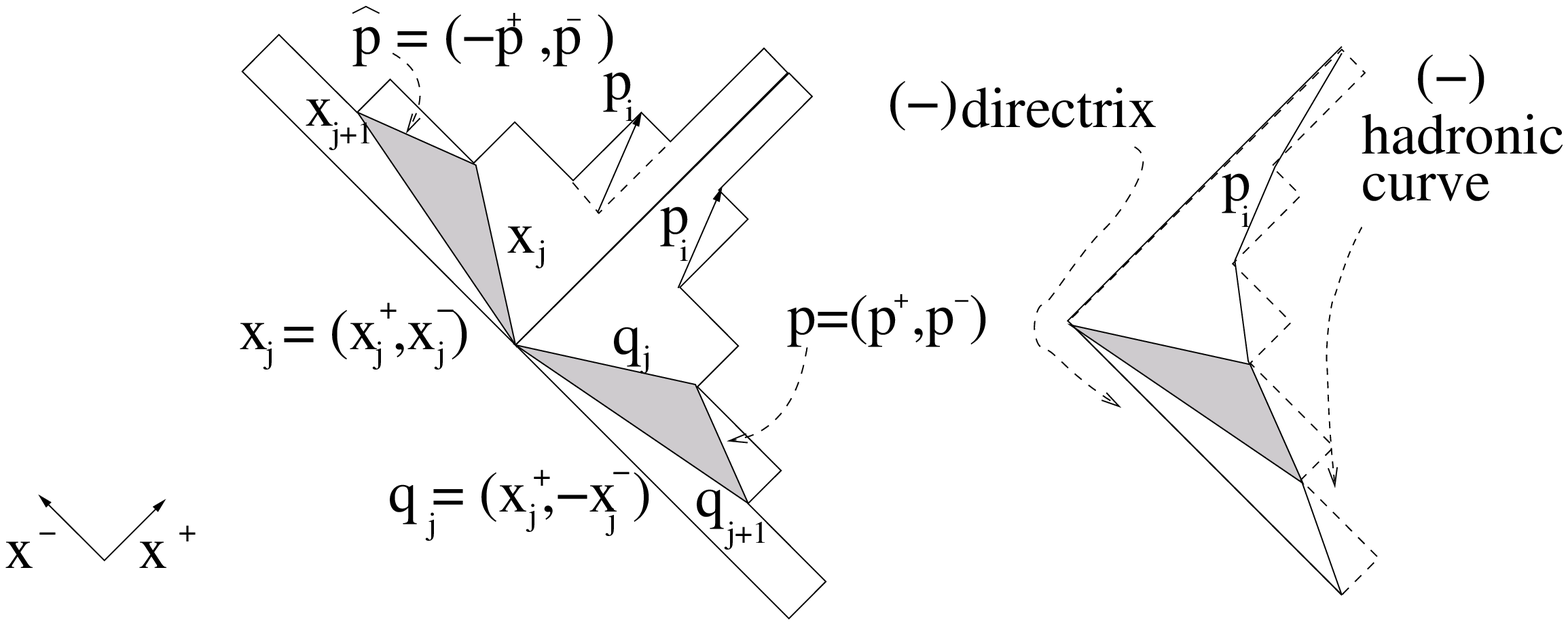, width=12.0cm }
\caption{One  possible way  to  partition the  area  of a  fragmenting
string  into contributions for  each hadron.  It shows  the connection
between the  area in the area law and the area  between the directrix
and the hadronic curve.}
\label{dual}
\end{center}
\end{figure}

It  is possible  to calculate  the  probability for  a string  break-up
inside the Lund  String Model using the basic  properties that we just
have  mentioned  if one  also  includes  a ''saturation  assumption'',
\cite{BA,BA1,BA2}.   This additional  assumption basically  means that
even if  the energy of the  original pair will  increase without limit
the distribution of  the proper times of the  production vertices will
stay finite. The two main results are:
\begin{itemize}
\item The probability  for
producing a hadron with mass $m$ taking a fraction $z$ of the remaining
(positive) light-cone momentum is given by $f(z)dz$, where
\begin{eqnarray}
\label{lundf}
f(z)&=& N\frac{1}{z}(1-z)^a exp(\frac{-bm^2}{z}) 
\end{eqnarray}
$f$  is called  the Lund  Symmetric Fragmentation  Function (probability
distribution).  We  note that small  $z$ values will  be exponentially
suppressed (with $\frac{1}{z}$), while  large $z$ values will be power
suppressed.  The final  result is that the vertices  typically will be
along a hyperbola in space-time in the Lund Model.

\item The Area Law, which states that the (non-normalised)
probability   to    produce   a   set   of    hadrons   with   momenta
$p_1,p_2,\ldots,p_n$ and total momentum $P_{tot}$ is given by:
\begin{eqnarray}
\label{arealaw}
dP_n(\{p_j\};P_{tot})=\prod_{j=1}^n N_j d^2p_j \delta({p_j}^2-{m_j}^2) \delta( 
\sum_{j=1}^n p_j - P_{tot}) \exp(-b  A)
\end{eqnarray}
, i.e. the  phase-space  times  a  negative  exponential  of  an
area.  The area  is the  area spanned  by the  string state  before it
decays, cf. Fig.\ref{area}.
\end{itemize}
\begin{figure}[t]
\begin{center}
\epsfig{figure=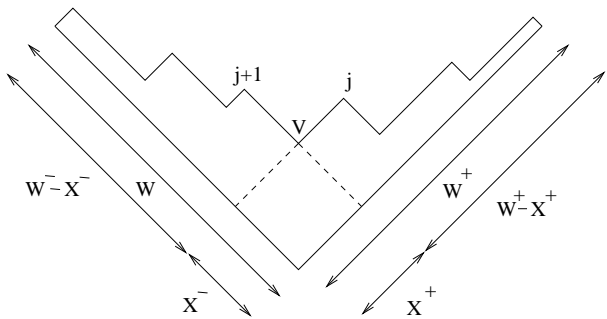, width=8.0cm }
\caption{The vertex V will subdivide the system into ''right-movers'' ($p_1,\ldots,p_j$) and 
''left-movers'' ($p_{j+1},\ldots,p_n$).}
\label{leftright}
\end{center}
\end{figure}
\begin{figure}[!h]
\mbox{
\hspace{0.5cm} \epsfig{file=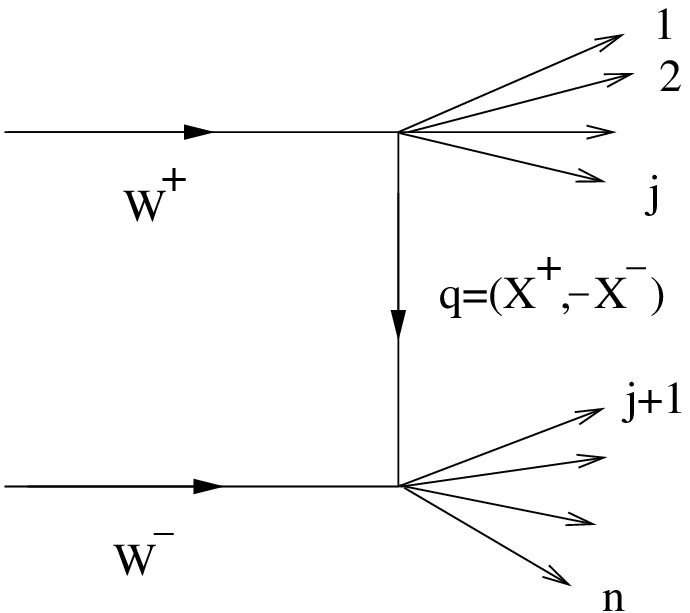,width=5.0cm}
\hspace{0.5cm} \epsfig{file=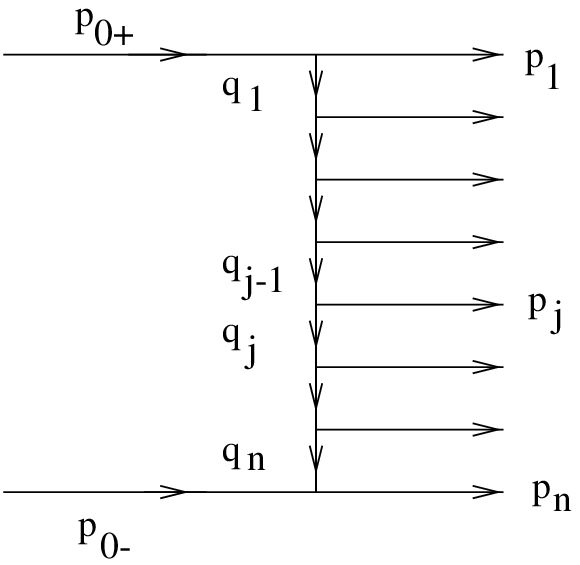,width=5.0cm}
}

\captive{The production process described in terms of momentum transfers.\label{ladderfig}}
\end{figure}
Up  to  this  point,  we   have  described  the  break-up  process  in
space-time. We  are now going to  show that one can  also describe the
same process  in momentum space.  In order to  do this, we  first note
that     a     vertex     (V)     subdivides    the     state     into
$\underbrace{\{p_n,\ldots,p_{j+1}\}}_{\mathrm{''to\,the\,left''}}\oplus\underbrace{\{p_{j},
\ldots,p_1\}}_{\mathrm{''to\,the\,right''}}$, cf. Fig.\ref{leftright}.
From this figure we conclude that:
\begin{eqnarray}
p_1+p_2+\ldots+p_j=(W^+-X^+,X^-) \\
\nonumber p_{j+1}+\ldots+p_n=(X^+,W^--X^-)
\end{eqnarray}
We  can therefore  describe  the process  according  to the  left-hand
figure in  Fig.\ref{ladderfig}, if we define the  momentum transfer $q
\equiv (X^+,-X^-)$.   This process can  be repeated for any  vertex.  This
means that String  Fragmentation can be described as  a ladder diagram
like the one in the right-hand figure of Fig.\ref{ladderfig}. There is
a  duality relationship  between  the space-time  picture  and this  last
diagram, cf. the left-hand figure of Fig.\ref{dual}.
 
\section{Fragmentation of Gluonic Strings}
\label{gfrag}

Presence  of internal gluonic  excitations cause  the string  to trace
complex  curved surfaces in  space--time.  Approximating  the partonic
state with a  finite number of gluons with  transverse momenta above a
certain scale is  equivalent to approximating the curved  surface of a
string  by a set  of planar  regions with  the (scaled)  parton energy
momenta  at the  boundaries. The  dynamics of  such a  string  is most
conveniently described in terms  of the action principle which implies
that the string world surface should be a minimal area surface.

This  means  that  the  details  of the  string  world  surface  are
completely determined  by a specification  of the boundary,  i.e., the
quark or  anti-quark world  lines. For an  open string these  two world
lines  differ by just  a translation,  and hence  it is  sufficient to
specify  only  one  of  them.   This  boundary  curve  is  called  the
directrix.  The  position of an  arbitrary point $X(t,\sigma)$  on the
surface  of  the string  (parametrised  by  time  '$t$' and  the  energy
$\sigma$ between the point and the  quark end point) is related to the
directrix ${\cal  A}(t)$ in the following way.
\begin{eqnarray}
\label{xsigmat}
X(t,\sigma)&=&\frac{1}{2}\left( {\cal  A}(t+\sigma)+{\cal  A}(t-\sigma)\right)
\end{eqnarray}
The function ${\cal A}(t)$ has the following periodicity and symmetry properties for a string which starts at a single point:
  
\begin{eqnarray}
\label{Aprop}
{\cal  A}(t+2E)&=&{\cal  A}(t)+2P \\
\nonumber {\cal  A}(-t)&=&-{\cal  A}(t)
\end{eqnarray}

We see from Eq.(\ref{xsigmat}) and Eq.(\ref{Aprop}) that the string surface is described by a
left moving and a right moving wave, which bounce at the endpoints. It
can also be  seen that the internal excitations  would move and affect
the end points  in colour order.  The directrix  of a string stretched
between a certain  set of partons is therefore  obtained by laying out
the  energy-momenta  of the  quark, the  gluons and  the  anti-quark in
colour order.

The minimal  area nature of the  string world surface  also means that
properties of  hadrons created during string  fragmentation are stable
against small  perturbations on the  boundary, i.e. that the  model is
infrared stable.

Since  the directrix describes  the string  state, hadronisation  of a
string  can  be  formulated  as  a process  along  the  directrix.  In
\cite{BASMFS1}   we   presented  one   such   formulation  of   string
fragmentation, and we briefly mention some of the main ideas here.

In \cite{BASMFS1}, we defined an iterative process along the directrix
to  describe string  fragmentation.  We  defined  (abstract) dynamical
variables  called  ``vertex''  vectors  $x_i$. These  variables  would
correspond  to  true  space-time  locations for  production  of  quark
anti-quark  pairs if  the  string had  no  gluonic excitations.   Each
particle  energy-momentum  vector $p_i$  is  constructed  as a  linear
combination of a vertex vector  $x_{i-1}$ and a certain section of the
directrix $k_i$.   The same pair  of vectors $x_{i-1}$ and  $k_i$ also
determines the next vertex vector $x_i$.

\begin{eqnarray}
\label{pjxjkj}
p_i=z_i   x_{i-1}+\frac{1}{2}(1-\frac{z_i^2   x_{i-1}^2}{m_i^2})   k_i \\
\nonumber x_i=(1-z_i)x_{i-1}+\frac{1}{2}(1+\frac{z_i^2 x_{i-1}^2}{m_i^2}) k_i
\end{eqnarray}

Since  with this construction,  $x_{i-1}+k_i=p_i+x_i$, we  associate a
``plaquette'' bounded by those  four vectors with the particle indexed
``i''. Each of the vectors $x_i$ is shared between two plaquettes, and
hence production of a set of  hadrons from a given string or directrix
means  construction  of  a  set  of  connected  plaquettes  along  the
directrix in  this model. The initial  value $x_0$, for  the series of
vertex vectors, is taken to be the first segment of the directrix. The
numbers $z_i$ are related to the area of the plaquettes.

There  is  a  one to  one  correspondence  between  the areas  of  the
plaquettes  constructed in  this way  and certain  areas which  can be
associated  with the production  of individual  hadrons in  the string
fragmentation model without gluonic excitations. Therefore, the sum of
the  areas of  the  plaquettes, which  is  also the  area between  the
directrix  and the  ``hadronic  curve'' (i.e.  the  curve obtained  by
placing  the  hadron  energy-momenta  one  after  the  other  in  rank
order. We called this curve the $X$-curve in \cite{BASMFS1}), could be
used as  the area in the Lund  Model area law. Therefore,  area law on
the whole  event can be  implemented by weighting the  construction of
each plaquette  according to its  area, which amounts to  choosing the
variables  ``$z_i$''  in   Eq.(\ref{pjxjkj})  according  to  the  Lund
symmetric fragmentation function.

It was  shown in  \cite{BASMFS1} that in  the limit  of infinitesimal
steps  $k_i$ along  the  directrix, the  hadronic  curve approaches  a
continuous  curve in  space--time which  we called  a  $P$-curve. This
could  be thought  of as  the equivalent  of the  average hyperbolical
fragmentation  region  in  a  simple gluon-less  string  fragmentation
scheme  in  this model.   There  is  a  close connection  between  the
$P$-curve  and a  certain  ${\cal X}$-curve defined  on the  partonic
state, or the directrix.  The  ${\cal X}$-curve and its length, called
the $\lambda$-measure  which is related  to the available  phase space
for hadron  production, are easily  calculated for a  given directrix.
It is apparent therefore that the average hadronic curve in this model
is determined by perturbative calculations.

\section{Generalised Dipoles}
\label{secgd}
In  this  section  we  will  discuss certain  properties  of  the  QCD
bremsstrahlung mostly  in connection with the dipole  cascade model as
implemented in  the Monte Carlo routine  {\Ar} and relate  them to the
hadronisation procedure we have just mentioned.

Gluon  radiation from  QCD colour  dipoles can  be compared  to photon
radiation from electromagnetic dipoles.  The inclusive density for the
emission of a gluon from a quark anti-quark dipole is given by

\begin{eqnarray}
\label{gdensity}
dn=\alpha_{\mathrm{eff}} \cdot \frac{dk_{\perp}^2} {k_{\perp}^2} \cdot dy \cdot
( \mathrm{Polarisation Sum})
\end{eqnarray}

This  is very  similar  to  the result  for  photons. Energy  momentum
conservation implies that the  transverse momentum and rapidity of the
emitted particle obey the following constraint:

\begin{eqnarray}
\label{cohcondn}
k_\perp cosh(y) \le \frac{\sqrt{s}}{2}
\end{eqnarray}

where $\sqrt{s}$ is  the center of mass energy of  the dipole. This in
turn, means that  the total rapidity range available  for the emission
is given by

\begin{eqnarray}
\label{phsp0g}
\Delta y \approx ln \left(\frac{s}{k_\perp^2}\right)
\end{eqnarray}

The emission of a  photon leaves the electromagnetic current unchanged
except  for small  recoil effects,  whereas the  emission of  a gluon,
which  has  colour  charge,  changes  the current  for  further  gluon
radiation.  However, it was shown in \cite{Azimov,DKMT} that to a very
good approximation, the density for  the emission of two gluons can be
written as the product of two contributions.

\begin{eqnarray}
\label{2gdensity}
dn(q,g_1,g_2,\bar{q})=dn(q,g_1,\bar{q})
\left[dn(q,g_2,g_1)+dn(g_1,g_2,\bar{q})-\mathrm{small \ correction}
\right]
\end{eqnarray}

Emission of two gluons could therefore be described as the emission of
one  first gluon  from the  original  quark anti-quark  pair, and  the
subsequent  emission  of  the   second  gluon  of  smaller  transverse
momentum, from  one or  the other colour  dipole involving  the doubly
(colour) charged  gluon.  In  other words, the  emission of  the first
gluon splits the original colour dipole into two dipoles which radiate
independently.

The available  rapidity range  for the emission  of the  second gluon,
which is  just the  sum of  the rapidity ranges  available in  the two
dipoles available after the first  emission, is more than the rapidity
range in the original quark anti-quark dipole.

\begin{eqnarray}
\label{phspincr}
\Delta y (g_2) = ln \left(\frac{s_{qg_1}}{k_\perp^2}\right)+ln \left(\frac{s_{g_1\bar{q}}}{k_\perp^2}\right) = ln \left(\frac{s}{k_\perp^2}\right) + ln \left(\frac{\frac{s_{qg_1}s_{g_1\bar{q}}}{s}}{k_\perp^2}\right)
\end{eqnarray}

The quantity $s_{q g_1}  s_{g_1 \bar{q}}/s$ is the invariantly
defined  transverse momentum  of the  first  gluon.  Eq.(\ref{phspincr})
therefore implies that  the emission of the first  gluon increases the
available  phase space  for further  gluon emission  by an  amount $ln
(k_{\perp 1}^2 / k_\perp^2)$.

\begin{figure}[t]
\begin{center}
\epsfig{figure=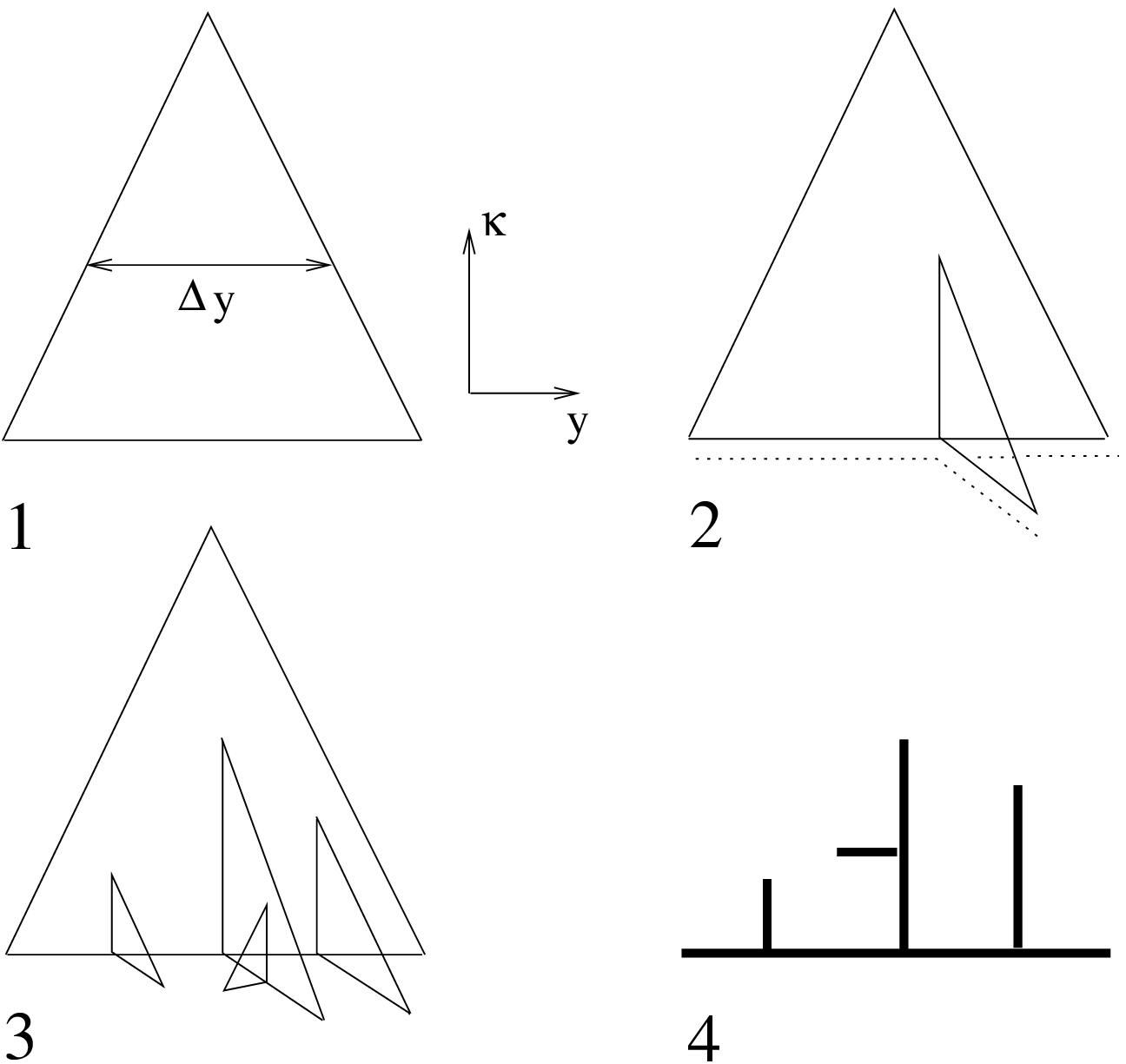, width=8.0cm }
\caption{The triangular Lund phase-space diagrams described in the text.}
\label{triangel}
\end{center}
\end{figure}

The  above dipole  splitting  scenario is  generalised  in the  dipole
cascade model. Emission of the second gluon is thought to give rise to
three dipoles  which radiate independently, and so  on.  The available
rapidity  range for  emission of  new gluons  at a  certain transverse
momentum scale $k_\perp$ is,  as a consequence of Eq.(\ref{phspincr}), a
function of all the emissions above that scale.

\begin{eqnarray}
\label{phspktsum}
\Delta y (g_n) = ln (\frac{s}{k_{\perp^2}}) + \sum_{i=1}^{n-1} ln (\frac{k_{\perp i}^2}{k_\perp^2})
\end{eqnarray}

The  increase of  phase space  due to  gluon emission  is conveniently
visualized  using the Lund  triangular phase  space diagrams,  such as
those  shown  in   Fig.\ref{triangel}.  The  $y$-axis  represents  the
logarithm of  the transverse momentum scale  $k_{\perp}^2$, denoted by
$\kappa$. Eq.(\ref{phsp0g})  then means that as  $\kappa$ is decreased
the available  phase space increases linearly with  $-\kappa$. This is
represented by the base line of the triangle which increases as $kappa$ decreases. In the second figure, we
show one gluon emitted at  a certain scale $\kappa_1$. The phase space
for  emission at  a  lower scale  is  the base  line  of the  original
triangle   plus   another   term   which   increases   linearly   with
$\kappa_1-\kappa$   according    to   Eq.(\ref{phspincr}).   This   is
represented  by  the   additional  triangular  appendage  starting  at
$\kappa_1$. The  total phase space is therefore  obtained by measuring
the  length  of the  base  line of  the  structure,  going around  all
triangular  appendages. The third  figure in  Fig.\ref{triangel} shows
this  after several emissions.  The forth  figure is  a view  from the
``bottom''  of the  phase space  triangle.  The  triangular appendages
appear as lines  coming out of the original  base line, increasing its
length.

The transverse  momentum scale $\kappa$  is a resolution scale  on the
partonic  state. The  smaller  the  scale, the  softer  the gluons  we
resolve.   Fig.\ref{triangel}  illustrates  that the  available  phase
space scales such that it increases faster than a linear rate when the
resolution scale is lowered. This  is a manifestation of the anomalous
dimensions of QCD.

It should be noted that the phase space in Eq.(\ref{phspktsum}) is not
written in terms of variables  describing the state from which further
emission  is  being considered.   It  requires  information about  the
history of the event, i.e.   which gluon was emitted first, which came
next etc., and  at what transverse momenta they  were emitted. It also
ignores  the effect of  recoil due  to gluon  emissions.  For  a small
invariant  mass  of the  radiating  dipole,  recoil  effects could  be
significant.  Emission from any one  dipole would give a recoil to the
two partons in that dipole, and would therefore change the mass of the
adjoining dipoles.   The assumption  of independent emission  from the
dipoles itself, however  provides one measure defined in  terms of the
dipoles available for radiation:

\begin{eqnarray}
\label{lmeasure}
\ell(k_\perp^2) = \sum^{\mathrm{dipoles}}
\ln \frac{s_{\mathrm{dipole}}}{k_\perp^2}
\end{eqnarray}

When a string  without gluons decays, the rapidity  range available to
hadrons  along  the  mean  hyperbolic  decay region  is  given  by  an
expression similar to  Eq.(\ref{phsp0g}), with the transverse momentum
scale  $k_\perp^2$, replaced by  a suitable  scale $m_0^2$  related to
hadronisation.  But  generalisation of  the phase space  available for
hadron formation  goes along  slightly different lines.   One observes
that the  hadronisation scale $m_0^2$  is independent of the  scale at
which the partonic state  is calculated. Since string fragmentation is
infrared stable,  the phase  space for hadrons  should be  an infrared
stable generalisation  of the flat  string phase space, as  softer and
softer  gluons are  added. Addition  of gluons  softer than  the scale
$m_0^2$ should not affect the phase space significantly.

Such a  generalisation was developed in \cite{BAGGBSII},  and has been
discussed over the years.  This measure, called the $\lambda$-measure,
is defined in terms of  differential equations for a general directrix
${\cal A}(\xi)$ as follows.  We define a four-vector valued functional
$q_T$ and a scalar valued functional $T$ :

\begin{eqnarray}
\label{Tint1}
T(\xi)=1+\int_0^{\xi}\frac{      d{\cal      A}(\xi_1)\cdot     d{\cal
A}(\xi_2)}{m_0^2}
\theta(\xi_1-\xi_2) T(\xi_2)
\end{eqnarray}

\begin{eqnarray}
\label{qTdef}
q_T(\xi)=         \frac{1}{T(\xi)}         \int_0^{\xi}         d{\cal
A}(\xi^{\prime})T(\xi^{\prime})
\end{eqnarray}
which satisfy the differential equations:
\begin{eqnarray}
\label{diffeq}
dT &=& \frac{q_Td{\cal  A}}{m_0^2} T \nonumber \\ dq_T  &=&d{\cal A} -
\frac{q_T      d{\cal     A}}{m_0^2}q_T      \nonumber\\     dq_T^2&=&
2(1-\frac{q_T^2}{m_0^2})q_Td{\cal A}
\end{eqnarray}
They can be integrated to:
\begin{eqnarray}
\label{Tint2}
T(\xi)&=&\exp\left(\int_0^{\xi}\frac{q_T(\xi^{\prime})d{\cal
      A}(\xi^{\prime})}{m_0^2}\right)  \nonumber   \\  q_T^2(\xi)  &=&
      m_0^2[1-T(\xi)^{-2}]
\end{eqnarray}

Finally  we note  that we  may  define a  four-vector valued  function
${\cal X}_{\mu}={\cal A}_{\mu}-q_{\mu}$ so that:
\begin{eqnarray}
\label{calXdef}
d{\cal X}=\frac{(q_Td{\cal A})}{m_0^2}q_T
\end{eqnarray}
Thus the vector  $q_T$  is the  tangent  to  the ${\cal  X}$-curve
reaching out to the  directrix. From Eq.(\ref{Tint2}) we conclude that
the functional $T$ is the exponential of an area (note that the vector
$d{\cal A}$ is light-like, and $qd{\cal A}$ is an area element between
the directrix and the ${\cal X}$-curve).  This area corresponds to the
measure  $\lambda$ mentioned  earlier. We  also note  that  the vector
$q_T$ quickly reaches the  length $m_0$.  Integrating the differential
equations for  a directrix built up  from a set  of light-like vectors
$\{k_j\}$ we obtain the iterative equations:
\begin{eqnarray}
\label{Tint3}
T_{j+1}&=&       T_j       (1+q_{T       j}k_{j+1}/m_0^2)       \equiv
\frac{T_j}{\gamma_{j+1}}
\nonumber \\
q_{T j+1}&=& \gamma_{j+1}q_{T j} + \frac{1+\gamma_{j+1}}{2}k_{j+1}
\end{eqnarray}
The starting values are $T_0=1$ and $q_{T 0}=k_0$.

The factorisability  of the functional  $T$ means that  its logarithm,
 $\lambda$,  can be  written as  a sum  containing one  term  for each
 $k_j$:
\begin{eqnarray}
\label{lambdapart}
\lambda \equiv \sum_{j=1}^{n+1} \Delta \lambda_j =
\sum_{j=1}^{n+1} \ln\left(1+q_{T j-1}k_{j}/m_0^2\right)
\end{eqnarray}

\begin{figure}
\begin{center}
\epsfig{figure=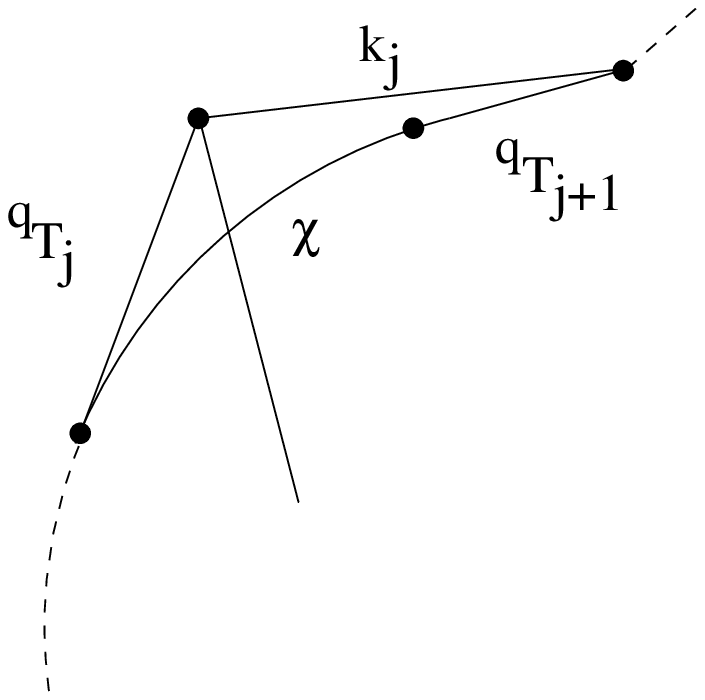, width=7.0cm }
\caption{A plaquette bordered by $q_{T j}$, $q_{T j+1}$
,  $k_j$ and  a  hyperbolic  segment from  the  ${\cal X}$-curve,  as
described in the text.}
\label{deltalambdaarea}
\end{center}
\end{figure}

The $\Delta \lambda_j$ defined in  this way corresponds to the size of
a subarea just as the  total $\lambda$ corresponds to the area spanned
between   the    ${\cal   X}$   and   the    directrix   curves.    In
Fig. \ref{deltalambdaarea}  we exhibit such a  region or ``plaquette''
bordered  by the ``initial''  $q_{T j}$  and the  ``final'' $q_{Tj+1}$
together with a  hyperbolic segment from the ${\cal  X}$-curve and the
gluon energy-momentum vector $k_j$ from the directrix.

Eq.(\ref{lambdapart})  should  be  compared with  Eq.(\ref{lmeasure}).
Eq.( \ref{lmeasure}) expresses the gluon emission phase space as a sum
of    contributions,    one    term    for   each    colour    dipole.
Eq.(\ref{lambdapart})  expresses  the  infrared  stable  hadronisation
phase space as a sum of  contributions, one term for each vector along
the  directrix, or one  term for  each plaquette  as mentioned  in the
previous  paragraph.   With  this  analogy  in mind  we  called  these
plaquettes   ``Generalised  Dipoles''(GD)   in   \cite{BASMFS2}.   The
hadronisation procedure outlined in  Sec. \ref{gfrag} could be thought
of as a procedure which divides these GDs into smaller plaquettes, one
for  each   hadron.  In   \cite{BASMFS2}  we  discussed   further  the
similarities  and differences between  the $\lambda$-measure,  and the
$\ell$-measure.

\begin{figure}
\begin{center}
\rotatebox{270}{\epsfig{figure=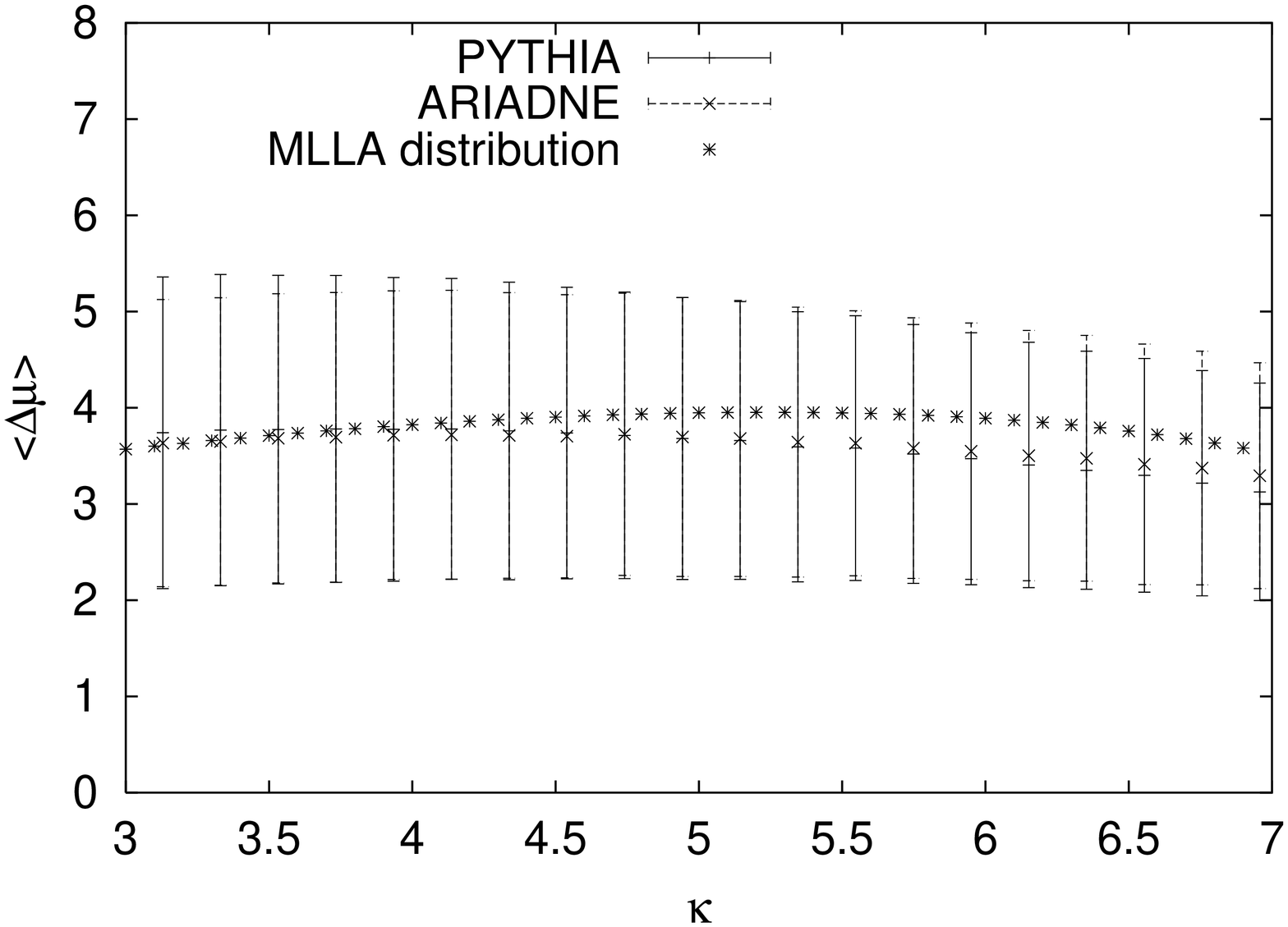, width=5.0cm }}\hspace{3mm}
\rotatebox{270}{\epsfig{figure=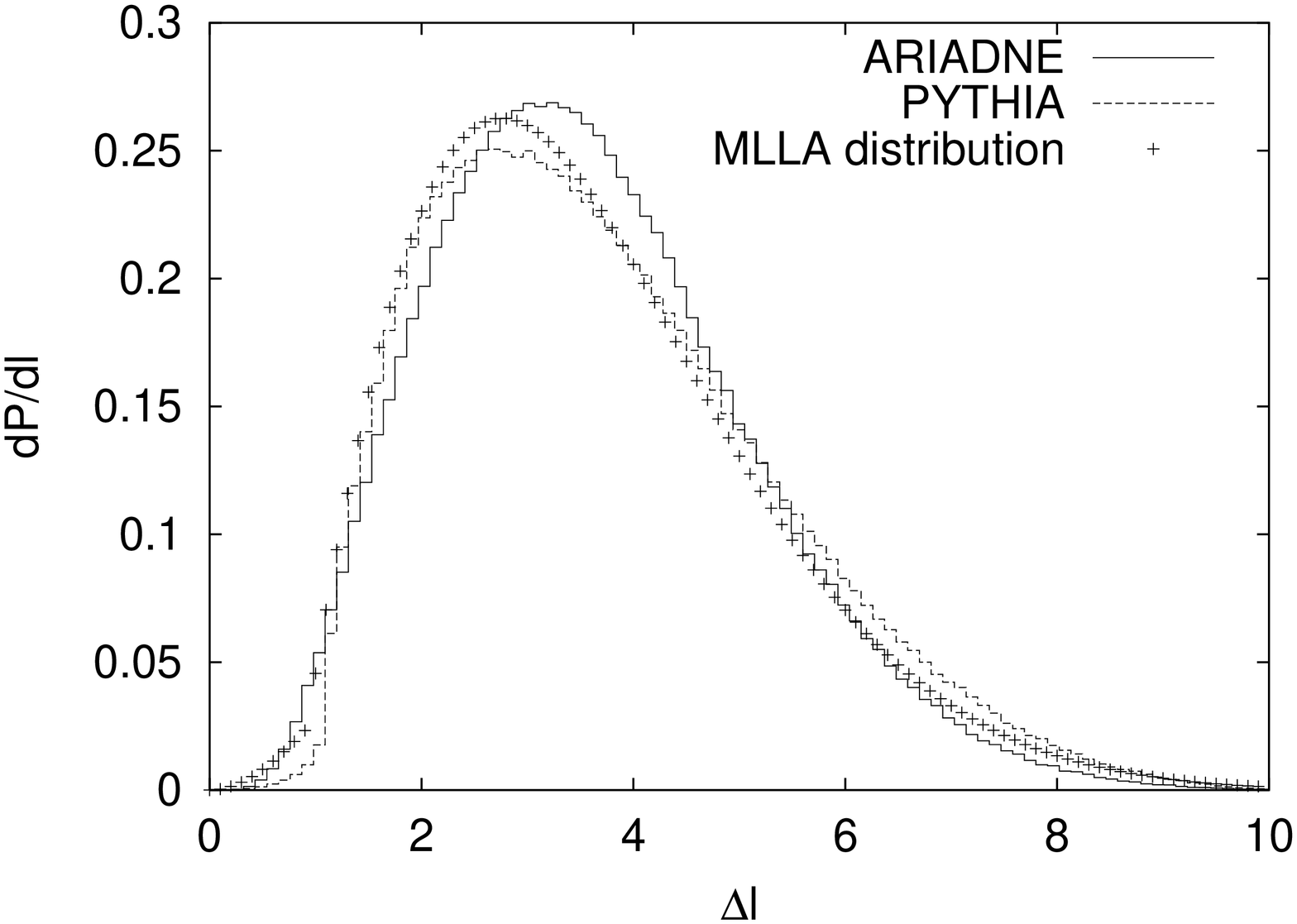, width=5.0cm }}
\caption{Dipole sizes obtained from \Py~ and \Ar~, measured relative to the cut-off scales, are relatively insensitive to the cut-off scales. The behaviour is quite well described by an MLLA calculation. }
\label{dipmlla}
\end{center}
\end{figure}

In \cite{BASMFS2}  we have also  reported on some  interesting scaling
properties of the contributions  to the $\ell$-measure from individual
dipoles. It  was found that  for partonic states obtained  from parton
showers  in  Monte Carlos  like  {\Py}  and  {\Ar}, the  dipole  sizes
measured relative  to the transverse momentum  cut-off scale $k_\perp$
had  a distribution which  was surprisingly  insensitive to  the scale
$k_\perp$ cf. Fig.\ref{dipmlla}.  We   showed  that  it   is  possible  to   understand  the
distributions in terms  of a ``gain-loss'' analysis: as  we go down in
transverse  momentum,  the  dipole  sizes  increase,  since  they  are
measured  relative  to transverse  momentum.  But  emission of  softer
gluons  would split some  dipoles of  a given  size into  smaller ones
which are  lost from a  certain bin in  dipole size. The  same process
would split  some dipoles of larger size,  so that they end  up in the
bin under consideration. An  integro differential equation taking into
account  these  considerations reproduces  the  observed behaviour  in
dipole sizes quite well, as illustrated in Fig.\ref{dipmlla}.

It was also found that the contributions to the $\lambda$-measure from
individual  GDs have  very similar  scaling properties,  if  the scale
$m_0^2$ is chosen  to be proportional to the  ordering variable in the
parton cascade.  Even  though there is no direct  way to calculate the
distribution of  sizes for GDs,  in the light of  certain similarities
between the  $\lambda$-measure and the  $\ell$-measure, such behaviour
is not completely unexpected.

That  parton showers  tend  to  produce structures  of  the same  size
relative   to  the   cut-off   scale  is   interesting  because   the
$\lambda$-measure and the  Generalised Dipoles are connections between
perturbative  physics and  hadronisation.  On  the one  hand  they are
related,  and  share  many  properties  with  phase  space  for  gluon
emission.   On the other  hand the  fragmentation process  reviewed in
Sec.\ref{gfrag} is  a connection between the plaquettes  we called GDs
and    the    plaquettes   we    defined    for    each   hadron    in
Sec.\ref{gfrag}. Properties of the  colour dipoles which are preserved
in the  Generalised Dipoles  will be carried  over to the  final state
hadrons.

\section{Conclusions}
String fragmentation can  be formulated as the production  of a set of
``plaquettes''  between  the  partonic  curve or  directrix,  and  the
hadronic curve. The sum of the areas of the plaquettes is taken as the
area in the Lund Model area law.  The mean size of this total area can
be  calculated   from  the   specified  perturbative  state,   and  is
proportional to the length of a curve called the ${\cal X}$-curve. The
length    of   the    ${\cal    X}$-curve   is    also   called    the
$\lambda$-measure. The ${\cal X}$-curve  looks like a set of connected
hyperbolae stretching across planar regions around the directrix which
we  called  ``Generalised Dipoles''.  The  size  of  the GDs  measured
relative  to  the  parton  shower   cut-off  scale  seems  to  have  a
distribution which  is insensitive to  the cut-off scale  itself. This
can be  understood as a balance  between emission of more  gluons at a
softer scale versus the scaling of the phase space variable itself.



\end{document}